\documentclass{jpsj-suppl}
\usepackage{graphicx}

\title{Hadron Spectroscopy in the Unquenched Quark Model\footnote{Talk presented at The 10th International Workshop on the Physics of Excited Nucleons (NSTAR2015), 
May 25-28 2015, Osaka (Japan).} }
\author{H. \textsc{Garc{\'{\i}}a-Tecocoatzi}$^{1,2}$, R. \textsc{Bijker} $^{1}$, J. \textsc{Ferretti}$^{3}$, E. \textsc{Santopinto}$^{2}$}
\inst{$^{1}$Instituto de Ciencias Nucleares, Universidad Nacional Aut\'onoma de M\'exico, AP 70-543, 04510 M\'exico DF, M\'exico \\
$^{2}$
INFN Sezione di Genova and Dipartimento di Fisica, Universit\`a di Genova, via Dodecaneso 33, I-16146 Genova, Italy\\
$^{3}$Dipartimento di Fisica and INFN, Universit\`a di Roma "Sapienza", Piazzale A. Moro 5, I-00185 Roma, Italy
}

\email{hugo.garcia@nucleares.unam.mx}

\recdate{September, 2015}

\abst{
We present the last applications  of the unquenched quark model (UQM)  to the  description of
 flavour asymmetry  and strangeness in the proton when baryon-meson components are included.    In the meson sector, the UQM is used to  
calculate the charmonium and bottomonium spectra with self-energy  
corrections due to the coupling to the meson-meson continuum. 
  }
 
\begin{document}

\maketitle

\section{Introduction}
The behavior of observables such as the spectrum and the magnetic moments of hadrons are well reproduced by   the   constituent quark model (CQM) \cite{Eichten:1974af,Isgur:1979be,Godfrey:1985xj,Capstick:1986bm,Giannini:2001kb,Glozman-Riska,Loring:2001kx,Ferretti:2011,Galata:2012xt,BIL}, even if  it neglects quark-antiquark pair-creation (or continuum-coupling) effects.
The unquenching of the quark model for hadrons is a way to take these components into account.

The unquenching of  CQM was initially   done by T\"ornqvist and collaborators, who used  an unitarized quark model \cite{Ono:1983rd,Tornqvist}, while 
Van Beveren and Rupp used a
t-matrix approach \cite{vanBeveren:1979bd,vanBeveren:1986ea}.
These techniques were applied to the  study of scalar meson nonet ($a_0$, $f_0$, etc.) of Ref. \cite{vanBeveren:1986ea,Tornqvist:1995kr} in which the loop contributions are given by the hadronic intermediate states that each meson can access. It is via these hadronic loops that the bare states become ``dressed'' and  the hadronic loop contributions totally dominate the dynamics of the process.  A similar approach was developed by Boglione and Pennington in Ref. \cite{Pennington:2002}, in which they investigated the dynamical generation of the scalar mesons by initially inserting only one ``bare seed''. On the other hand, Isgur and coworkers in Ref. \cite{Geiger:1989yc} demonstrated that the effects of the $q \bar q$ sea pairs in meson spectroscopy is simply a renormalization of the meson string tension.  The  
strangeness content of the nucleon and electromagnetic form factors  
were also investigated, see refs. \cite{Geiger:1996re,Bijker:2012zza}, whereas  Capstick and Morel in Ref. \cite{Capstick} analyzed  
baryon meson loop effects on the spectrum of nonstrange baryons.   
In the meson sector, Eichten {\it et al.} explored the influence of the open-charm channels on the charmonium properties using the Cornell coupled-channel model \cite{Eichten:1974af} to assess departures from the single-channel  potential-model expectations.

In this work we present  the latest applications of the UQM to study the flavor asymmetry  and strangeness  of the proton,
 in wich the effects of the sea quarks were introduced  into the CQM in a systematic way and the wave fuctions were given explicitly.  Finally, the UQM is applied 
 to describe meson observables and the spectroscopy of the charmonium  and bottomonium.	

\section{The Unquenched Quark Model }
\label{Sec:formalism}
In the UQM for baryons \cite{Bijker:2012zza,Santopinto:2010zza,Bijker:2009up,Bijker:210} and mesons \cite{bottomonium,charmonium,Ferretti:2013vua,Ferretti:2014xqa}, the hadron wave function is made up of a zeroth order $qqq$ ($q \bar q$) configuration plus a sum over the possible higher Fock components, due to the creation of $^{3}P_0$ $q \bar q$ pairs. Thus,  we have 
\begin{eqnarray} 
	\label{eqn:Psi-A}
	\mid \psi_A \rangle
	 ={\cal N} \left[ \mid A \rangle 
	+ \sum_{BC \ell J} \int d \vec{K} \, k^2 dk \, \mid BC \ell J;\vec{K} k \rangle \right.
	\left.  \frac{ \langle BC \ell J;\vec{K} k \mid T^{\dagger} \mid A \rangle } 
	{E_a - E_b - E_c} \right] ~, 
\end{eqnarray}
where $T^{\dagger}$ stands for the $^{3}P_0$ quark-antiquark pair-creation operator \cite{bottomonium,charmonium,Ferretti:2013vua,Ferretti:2014xqa}, $A$ is the baryon/meson, $B$ and $C$ represent the intermediate state hadrons. $E_a$, $E_b$ and $E_c$ are the corresponding energies, $k$ and $\ell$ the relative radial momentum and orbital angular momentum between $B$ and $C$ and $\vec{J} = \vec{J}_b + \vec{J}_c + \vec{\ell}$ is the total angular momentum. 
It is worthwhile noting that in Refs. \cite{bottomonium,charmonium,Ferretti:2013vua,Ferretti:2014xqa,Kalashnikova:2005ui}, the constant pair-creation strength in the operator (\ref{eqn:Psi-A}) was substituted with an effective one, to suppress unphysical heavy quark pair-creation. 

 
The introduction of continuum effects in the CQM can thus be essential to study observables that only depend on $q \bar q$ sea pairs, like the strangeness content of the nucleon electromagnetic form factors \cite{Geiger:1996re,Bijker:2012zza} or the flavor asymmetry of the nucleon sea \cite{Santopinto:2010zza}. 
The  continuum effects can give  important corrections to baryon/meson observables, like the self-energy corrections to meson masses \cite{bottomonium,charmonium,Ferretti:2013vua,Ferretti:2014xqa} or the importance of the orbital angular momentum in the spin of the proton \cite{Bijker:2009up}.

\section{Flavour content in the proton}
 
The  evidence for the flavor asymmetry of the proton sea was found 
by NMC at CERN \cite{nmc}. The flavor asymmetry in the proton is related to the Gottfried integral 
for the difference of the proton and neutron electromagnetic structure functions 
\begin{eqnarray}
S_G = \int_0^1 dx \frac{F_2^p(x)-F_2^n(x)}{x} 
= \frac{1}{3} - \frac{2}{3} \int_0^1 dx \left[ \bar{d}(x) - \bar{u}(x) \right] ~.
\end{eqnarray}
If one  takes  a flavor symmetric sea, one obtains the 
Gottfried sum rule $S_G=1/3$, but the final NMC value is $0.2281 \pm 0.0065$ at $Q^2 = 4$ 
(GeV/c)$^2$ for the Gottfried integral over the range $0.004 \leq x \leq 0.8$ \cite{nmc}, 
which implies a flavor asymmetric sea. The Gottfried sum rule has been 
confirmed by other experimental collaborations \cite{hermes,nusea}. 
Theoretically, it was shown  in Ref. \cite{Thomas}, that the coupling of the nucleon to the pion 
cloud provides a natural mechanism to produce a flavor asymmetry. 
\begin{figure}
\begin{center}
\includegraphics[width=6cm]{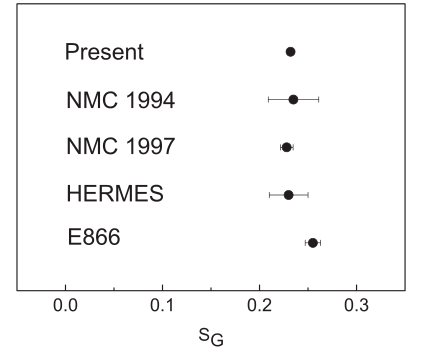}
\caption{\label{protonasy}Comparison the value of Gottfried sum rule calculated within UQM with the experimental data  from NMC 1994, NMC 1997, HERMES, and E866. Figure  taken 
from Ref. \cite{Bijker:210}; APS copyright. }
\end{center}
\end{figure}
In the UQM, the flavor asymmetry can be 
calculated from the difference of the probability to find  $\bar{d}$ and $\bar{u}$ sea quarks in the proton. Our result is shown in  Fig. \ref{protonasy}.

\begin{figure}[h]
\begin{minipage}{18pc}
\includegraphics[width=18pc]{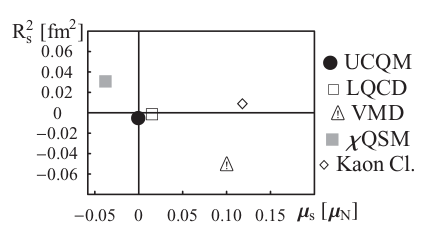}
\caption{\label{smagnetic}TheUQM results for the strange magnetic moment and radius of the proton. Figure taken from Ref. \cite{Bijker:2012zza}; APS copyright.}
\end{minipage}\hspace{2pc}%
\begin{minipage}{15pc}
\includegraphics[width=15pc]{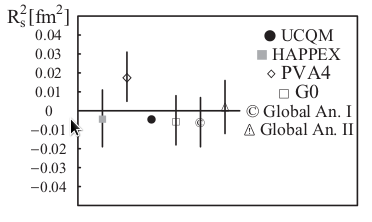}
\caption{\label{sradio} Comparison between our resulting value for the strange radius of the proton in the UQM. Figure  taken from Ref. \cite{Bijker:2012zza};  APS copyright.}
\end{minipage} 
\end{figure}

In a second stage, we calculated the strangeness content of the nucleon, see ref. \cite{Bijker:2012zza}. In the UQM the strange magnetic moment of the proton is defined as the expectation value of the operator
\begin{eqnarray}
\vec{\mu}_{s}=\sum_{i} \mu_{i,s}\left[2\vec{s}(q_{i})+\vec{l}(q_{i})- 2\vec{s}(\bar{q}_{i})-\vec{l}(\bar{q}_{i}) \right]
\end{eqnarray}
on the proton state of Eq. (\ref{eqn:Psi-A}), which represents the contribution of the strange quarks to the magnetic moment of the proton; $\mu_{i,s}$ is the magnetic moment of the quark i times a projector on strangeness and the strange quark magnetic moment is set as in Ref. \cite{Bijker:210}. Our result  is  $\vec{\mu}_{s}=0.0006 \mu_N$(see Fig.\ref{smagnetic}).

Similarly, the strange radius of the proton is defined as the expectation value of the operator
\begin{eqnarray}
R^{2}_{s}= \sum^{5}_{i=1}e_{i,s}\left(\vec{r}_{i}-\vec{R}_{\rm cm}\right)^2
\end{eqnarray}
on the proton state of Eq. (\ref{eqn:Psi-A}), where $e_{i,s}$ is the electric charge
of the quark $i$ times a projector on strangeness, $\vec{r}_{i}$  and $\vec{R}_{\rm cm}$ are the coordinates of the quark $i$ and of the intermediate state center of mass, respectively.
The expectation value of $R^{2}_{s}$ on the proton is equal to $-0.004 {\rm fm}^2 $. In Fig. \ref{sradio} our result is compared with the experimental data.

\section{Self-energy corrections in the UQM}
The method was used by some of us to compute the charmonium  ($c \bar c$) and
bottomonium ($b \bar b$) spectra with self-energy corrections, due to continuum coupling effects  \cite{bottomonium,charmonium,Ferretti:2013vua,Ferretti:2014xqa}. 
In the UQM, the physical mass of a meson, 
\begin{equation}
	\label{eqn:self-trascendental}
	M_a = E_a + \Sigma(E_a)  \mbox{ },
\end{equation}
is given by the sum of two terms: a bare energy, $E_a$, calculated within a potential model \cite{Godfrey:1985xj}, and a self energy correction, 
\begin{equation}
	\label{eqn:self-a}
	\Sigma(E_a) = \sum_{BC\ell J} \int_0^{\infty} k^2 dk \mbox{ } \frac{\left| M_{A \rightarrow BC}(k) \right|^2}{E_a - E_b - E_c}  \mbox{ },
\end{equation}
computed within the UQM formalism. 

\begin{figure}[h]
\begin{minipage}{16pc}
\includegraphics[width=16pc]{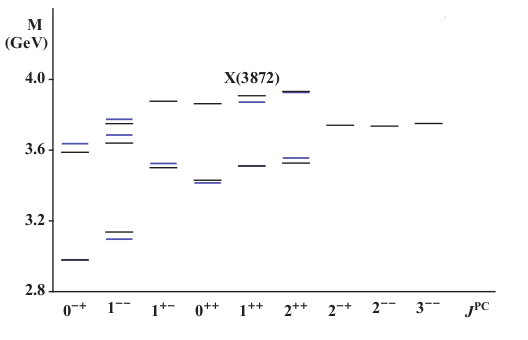}
 \caption{\label{charm} Charmonium spectrum with self energies corrections. 
  Black lines are theoretical predictions and blue lines are experimental data available. Figure taken from Ref. \cite{charmonium}; APS copyright.}\end{minipage}\hspace{2pc}%
\begin{minipage}{18pc}
\includegraphics[width=18pc]{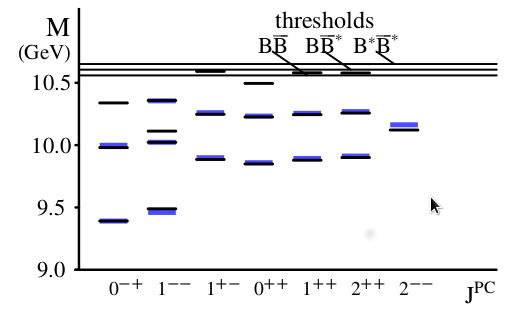}
 \caption{\label{botton}Bottomonium spectrum with self energies corrections.
 Black lines are theoretical predictions and blue lines are experimental data available. Figure taken from Ref. \cite{Ferretti:2013vua}; APS copyright. 
 }

\end{minipage} 
\end{figure}
Our results for the self energies corrections  of charmonia \cite{charmonium,Ferretti:2014xqa} and bottomonia \cite{bottomonium,Ferretti:2013vua,Ferretti:2014xqa} spectrums, 
are shown in  figures \ref{charm} and \ref{botton}.

\section{Discussion and conclusion}
The asymmetry and "strangeness" observables in the proton can only be understood  when continuum components in the wave function are included.
Our results, as shown in  Figures  \ref{protonasy}, \ref{smagnetic}  and  \ref{sradio}, are in agreement with the experimental data.

The self energies corrections  of charmonium  and bottomonium spectra, see figures \ref{charm} and \ref{botton}, show that the pair-creation effects on the spectrum of heavy mesons are quite small. Specifically for charmonium and bottomonium states, they are of the order of $2 - 6\%$ and $1 \%$, respectively. 
The relative mass shifts, i.e. the difference between the self energies of two meson states, are in the order of a few tens of MeV. 

In our framework  the $X(3872)$ can be interpreted as a $c \bar c$ core [the $\chi_{c1}(2^3P_1)$], plus higher Fock components due to the coupling to the meson-meson continuum. In Ref. \cite{Ferretti:2014xqa}, we obtained   that the probability to find the $X(3872)$ in its core or continuum components is approximately $45\%$ and $55\%$, respectively.  

In conclusion, the flavor asymmetry in the  proton is  well described by the UQM. The "strangeness" observables of the proton are found to be  
negligible and our results compatible with the latest experimental  
data and recent lattice calculations. Finally, our self energies corrections  for 
 charmonium  and bottomonium spectra are found to be significant.

\section*{Acknowledgments}

This work is supported in part by PAPIIT-DGAPA, Mexico (grant IN107314) and INFN sezione di Genova .


\begin{thebibliography}{9}

  
\bibitem{Eichten:1974af}  
  ~E. Eichten, K. ~Gottfried, T. ~Kinoshita , J.~B.~Kogut,  K. ~D. ~Lane  and T.  ~-M. ~Yan: 
 {\it Phys.\ Rev.\ Lett.\ }  {\bf 34} (1975) 369;
  ~E. Eichten, K. ~Gottfried, T. ~Kinoshita , K.  ~D. ~Lane  and T.  ~-M. ~Yan:
  {\it  Phys.\ Rev.} \ D {\bf 17} (1978) 3090;  {\it  Phys.\ Rev.} \ D
  {\bf 21} (1980)  203. 
	
\bibitem{Isgur:1979be}
N. ~Isgur and  G. ~Karl:
   {\it Phys.\ Rev.}\  D {\bf 18} (1978) 4187;
    {\it Phys.\ Rev.}\  D
  {\bf 19} (1979)  2653 
  [ Erratum-ibid.\  {\bf 23} (1981) 817];
  {\it Phys.\ Rev.}\  D  {\bf 20} (1979) 1191.
	
\bibitem{Godfrey:1985xj}
 S. ~Godfrey  and N. ~Isgur:
  {\it Phys.\ Rev.}\ D {\bf 32} (1985) 189.

\bibitem{Capstick:1986bm}
S.  ~Capstick  and N. ~Isgur: 
   {\it Phys.\ Rev.}\  D {\bf 34} (1986) 2809 .
	
\bibitem{Giannini:2001kb}
 M. ~Ferraris, M. M.  ~Giannini, M. ~Pizzo, E. ~Santopinto  and L. ~Tiator:
  {\it Phys.\ Lett.} \ B {\bf 364} (1995) 231; 
 E. ~Santopinto, F. ~Iachello  and M. M. ~Giannini:
   {\it Eur.\ Phys.\ J.} \ A {\bf 1} (1998) 307 ;
E. ~ Santopinto   and G. ~Giannini:
   {\it Phys.\ Rev.}\ C {\bf 86} (2012) 065202;	
   M. M. Giannini, E. Santopinto:  {\it Chin. J. Phys.}  {\bf 53}  (2015) 020301; 
 M. Aiello,  {\it et al.: Phys.Lett.} B (1996) 387, 215; Aiello {\it et al.}  {\it
 J. of Phys.} G  {\bf 24} (1988) 753; R. Bijker,  F.  Iachello,  E. Santopinto:
 {\it J. of  Phys.} A {\bf 31} (1988) 9041; M. De Sanctis  {\it et al. :} {\it Phys. Rev.}  C {\bf 6} (2007)
062201; MM.Giannini {\it et al.}: {\it Eur. Phys. J.} A {\bf 12 } (2001)447;
De Sanctis et al.:    {\it Phys.\ Rev.}\  C {\bf 62} (2000) 025208;
 M. De Sanctis et al.:
{\it Eur. Phys. J.} A {\bf 1 }, (1998) 187; De Sanctis, M. M. Giannini, E. Santopinto {\it et al.}: 
{\it Nucl. Phys.} A {\bf 755} (2005) 294; De Sanctis, M. M. Giannini, E. Santopinto {\it et al.}: 
{\it Eur. Phys. J.} A {\it 19} (2005) 81; E. Santopinto, A. Vassallo, M. M. Giannini
{\it et al.}: 	 {\it Phys.\ Rev.}\  C {\bf 82} (2010) 065204.
\bibitem{Glozman-Riska}
L. Y.   ~Glozman, D. O. ~Riska: 
   {\it Phys.\ Rept.}\  {\bf 268} (1996) 263;
 L. Y. ~Glozman, W. ~Plessas, K. ~Varga, R. F. ~Wagenbrunn:
   {\it Phys.\ Rev.}\  D {\bf 58} (1998) 094030.  
  
\bibitem{Loring:2001kx}
U. ~Loring, B. C.  ~Metsch and H. R. ~Petry:
  {\it Eur.\ Phys.\ J.} \  A {\bf 10} (2001) 395. 
	
\bibitem{Ferretti:2011}
E.  ~Santopinto: 
   {\it Phys.\ Rev.}\  C {\bf 72} (2005) 022201;
J.  ~Ferretti, A. ~Vassallo and E. ~Santopinto:
   {\it Phys.\ Rev.}\ C {\bf 83} (2011) 065204;
  M. ~De~Sanctis, J.  ~Ferretti, E. ~Santopinto   and A. ~Vassallo: 
  arXiv:1410.0590; M. De Sanctis,  J. Ferretti,
E. Santopinto: {\it  Phys. Rev. } C {\bf 84} (2011) 055201;
E. Santopinto; J. Ferretti: Phys. Rev. C 92 ,
  025202( 2015 )

	
\bibitem{Galata:2012xt}
G. ~Galat\`a and E. ~Santopinto:
   {\it Phys.\ Rev.}\ C {\bf 86} (2012) 045202.	
		
\bibitem{BIL}
 R. Bijker, F. Iachello  and  A. Leviatan:
 {\it Ann. Phys.} (N.Y.) {\bf 236} (1994) 69; 
 {\it Phys. Rev.} C {\bf 54} (1996) 1935; 
 {\it Phys. Rev. }D {\bf 55} (1997) 2862; 
 {\it Ann. Phys. } (N.Y.) {\bf 284} (2000) 89.
	
	
\bibitem{vanBeveren:1979bd} 
  E. ~van Beveren,  C. ~Dullemond and G. ~Rupp:
   {\it Phys.\ Rev. }\ D {\bf 21} (1980) 772
  [  Erratum-ibid.\ D {\bf 22} (1980) 787].

\bibitem{vanBeveren:1986ea} 
 E ~van Beveren, T.  A. ~Rijken, K. ~Metzger, C. ~Dullemond, G.  ~Rupp  and J. E. ~Ribeiro: 
   {\it Z.\ Phys.}\ C {\bf 30} (1986) 615.
  \bibitem{Ono:1983rd}
S.  ~Ono  and N.  A. ~T\"ornqvist:
   {\it Z.\ Phys.}\ C {\bf 23} (1984) 59;
  K. ~Heikkila, S. ~Ono  and N. A.  ~T\"ornqvist: 
   {\it Phys.\ Rev.} \ D {\bf 29} (1984) 110
  [ Erratum-ibid.\ {\bf 29} (1984) 2136];
 S.  ~Ono, A. I ~Sanda  and N.  A. ~T\"ornqvist: 
   {\it Phys.\ Rev.}\ D {\bf 34} (1986) 186.  
		
\bibitem{Tornqvist}
  N.~A. ~T\"ornqvist and  P. ~Zenczykowski: 
   {\it Phys.\ Rev.}\  D {\bf 29} (1984) 2139; 
   {\it Z.\ Phys.}\  C {\bf 30} (1986) 83;
  P. ~Zenczykowski: 
   {\it Annals Phys.}\  {\bf 169} (1986) 453.	
	
\bibitem{Tornqvist:1995kr} 
N. A. ~Tornqvist: 
  {\it Z.\ Phys.}\ C {\bf 68} (1995) 647.
\bibitem{Pennington:2002} 
  M. ~Boglione  and M. R.  ~Pennington: 
     {\it Phys.\ Rev.}\ D {\bf 65} ( 2002) 114010.	

	
\bibitem{Geiger:1989yc} 
P. ~Geiger and N.  ~Isgur: 
     {\it Phys.\ Rev.}\ D {\bf 41} (1990) 1595.
	
\bibitem{Geiger:1996re} 
P. ~Geiger   and N.  ~Isgur:
     {\it Phys.\ Rev.}\ D {\bf 55} (1997) 299.	
		
\bibitem{Bijker:2012zza}  
  R. ~Bijker, J. ~Ferretti  and  E. ~Santopinto: 
     {\it Phys.\ Rev.}\ C {\bf 85} (2012) 035204.   
\bibitem{Capstick}
S. Capstick and  D. Morel:  [nucl-th/0204014].
  
\bibitem{Santopinto:2010zza}
E ~Santopinto   and R. ~Bijker: 
   {\it  Phys.\ Rev.}\ C {\bf 82} (2010) 062202.     
  
\bibitem{Bijker:2009up}
  E. ~Santopinto  and R. ~Bijker: 
    {\it Few Body Syst.}\  {\bf 44} (2008) 95.
\bibitem{Bijker:210}  
  R. ~Bijker and E. ~Santopinto:  
     {\it Phys.\ Rev.}\ C {\bf 80} (2009) 065210. 
		
\bibitem{bottomonium}
  J. ~Ferretti, G. ~Galat\`a, E. ~Santopinto and A. ~Vassallo: 
     {\it Phys.\ Rev.} \ C {\bf 86} (2012) 015204.  
	
\bibitem{charmonium}  
 J. ~Ferretti,  G.  ~Galat\`a  and E. ~Santopinto: 
    {\it Phys.\ Rev.}\ C {\bf 88} (2013) 015207.	
	
\bibitem{Ferretti:2013vua}   
J. ~Ferretti,  and E. ~Santopinto:  
     {\it Phys.\ Rev.}\  D {\bf 90} (2014) 094022.	
  
\bibitem{Ferretti:2014xqa} 
J. ~Ferretti, G. ~Galat\`a   and  E. ~Santopinto: 
     {\it Phys.\ Rev.} \  D {\bf 90} (2014) 054010.	
\bibitem{Kalashnikova:2005ui}
Y. S. ~Kalashnikova:  
    {\it Phys.\ Rev.}\ D {\bf 72} (2005) 034010.		
\bibitem{nmc}
 P.  Amaudruz,  {\it et al.}:  {\it Phys. Rev. Lett.} {\bf 66}  (1991) 2712; 
M. Arneodo {\it et al.}:  {\it Nucl. Phys. } B {\bf 487} (1997) 3.

\bibitem{hermes}
 K. Ackerstaff {\it et al.}:  {\it Phys. Rev. Lett.} {\bf 81} (1988) 5519. 

\bibitem{nusea}
R. S. Towell  {\it et al.}:  {\it Phys. Rev. }D {\bf 64} (2001) 052002.

\bibitem{Thomas}
A. W. Thomas:   {\it Phys. Lett. }B {\bf 126}  (1983) 97.

\end{thebibliography}
\end{document}